\documentclass[fleqn,usenatbib,useAMS]{mnras}
 \usepackage{graphicx}
 \usepackage{amsmath}
 \usepackage{amssymb}
 \usepackage[T1]{fontenc}
 \usepackage{ae,aecompl}
 \usepackage{txfonts}
 \usepackage{enumerate}

 \title[Dynamical evolution of 1991~VG]
       {Dynamical evolution of near-Earth asteroid 1991~VG}

 \author[C. de la Fuente Marcos and R. de la Fuente Marcos]
        {C.~de~la~Fuente~Marcos\thanks{E-mail: nbplanet@ucm.es}
         and
         R. de la Fuente Marcos \\
         Universidad Complutense de Madrid,
         Ciudad Universitaria, E-28040 Madrid, Spain}
 \date{Accepted 2017 September 26. 
       Received 2017 September 26; 
       in original form 2017 August 4}
 \pubyear{2017}
 
\begin{document}
  \label{firstpage}
  \pagerange{\pageref{firstpage}--\pageref{lastpage}}
  \maketitle

  \begin{abstract}
     The discovery of 1991~VG on 1991 November 6 attracted an unprecedented 
     amount of attention as it was the first near-Earth object (NEO) ever 
     found on an Earth-like orbit. At that time, it was considered by some as 
     the first representative of a new dynamical class of asteroids, while 
     others argued that an artificial (terrestrial or extraterrestrial) 
     origin was more likely. Over a quarter of a century later, this peculiar 
     NEO has been recently recovered and the new data may help in confirming 
     or ruling out early theories about its origin. Here, we use the latest 
     data to perform an independent assessment of its current dynamical 
     status and short-term orbital evolution. Extensive $N$-body simulations 
     show that its orbit is chaotic on time-scales longer than a few decades. 
     We confirm that 1991~VG was briefly captured by Earth's gravity as a 
     minimoon during its previous fly-by in 1991--1992; although it has been 
     a recurrent transient co-orbital of the horseshoe type in the past and 
     it will return as such in the future, it is not a present-day co-orbital 
     companion of the Earth. A realistic NEO orbital model predicts that 
     objects like 1991~VG must exist and, consistently, we have found three 
     other NEOs ---2001~GP$_{2}$, 2008~UA$_{202}$ and 2014~WA$_{366}$--- 
     which are dynamically similar to 1991~VG. All this evidence confirms 
     that there is no compelling reason to believe that 1991~VG is not 
     natural.
  \end{abstract}


  \section{Introduction}
     Very few near-Earth objects (NEOs) as small as 1991~VG (about 10~m) have given rise to so much controversy and imaginative conjectures.
     Asteroid 1991~VG was the first NEO ever discovered moving in an orbit that is similar to that of the Earth (Scotti \& Marsden 1991). 
     This element of novelty led to a stimulating debate on how best to interpret the new finding. On the one hand, it could be the first 
     member of a new orbital class of NEOs (e.g. Rabinowitz et al. 1993); on the other, it could be a relic of space exploration (e.g. 
     Scotti \& Marsden 1991; Steel 1995b). 

     In addition to the primary debate on the possible character ---natural versus artificial--- of 1991~VG, a lively discussion resulted in 
     multiple theories about its most plausible origin; the main asteroid belt (e.g. Brasser \& Wiegert 2008), lunar ejecta (e.g. Tancredi 
     1997), a returning spacecraft (e.g. Steel 1995b) or space debris (e.g. Scotti \& Marsden 1991), and being an extraterrestrial artefact 
     (e.g. Steel 1995a), were all argued in favour and against as possible provenances for this object. After being last observed in 1992 
     April, 1991~VG has spent over a quarter of a century at small solar elongation angles, out of reach of ground-based telescopes.

     Now, this peculiar minor body has been recovered (Hainaut, Koschny \& Micheli 2017)\footnote{\url{http://www.minorplanetcenter.net/mpec/K17/K17L02.html}} 
     and the new data may help in confirming or ruling out early speculative theories about its origin. Here, we use the latest data 
     available on 1991~VG to study its past, present and future orbital evolution in an attempt to understand its origin and current 
     dynamical status. This paper is organized as follows. In Section 2, we present historical information, current data, and the various 
     theories proposed to explain the origin of 1991~VG. Details of our numerical model and 1991~VG's orbital evolution are presented and 
     discussed in Section 3. In Section 4, we show that even if 1991~VG is certainly unusual, other known NEOs move in similar orbits and 
     orbital models predict that such objects must exist naturally. Arguments against 1991~VG being a relic of alien or even human space
     exploration are presented in Section 5. Our results are discussed in Section 6. Section 7 summarizes our conclusions.    

  \section{Asteroid 1991~VG: data and theories}
     Asteroid 1991~VG was discovered on 1991 November 6 by J.~V. Scotti observing with the Spacewatch 0.91-m telescope at Steward 
     Observatory on Kitt Peak at an apparent visual magnitude of 20.7, nearly 0.022~au from the Earth (Scotti \& Marsden 1991). The rather 
     Earth-like orbit determination initially led to suspect that the object was of artificial origin, i.e. returning space debris, perhaps 
     a Saturn S-IVB third stage. It experienced a close encounter with our planet at nearly 0.0031~au on 1991 December 5 and with the Moon 
     at 0.0025~au on the following day, but it was not detected by radar at NASA's Goldstone Deep Space Network on December 12 (Scotti et 
     al. 1991). 

     After being last imaged in 1992 April, 1991~VG remained unobserved until it was recovered on 2017 May 30 by Hainaut et al. (2017) 
     observing with the Very Large Telescope (8.2-m) from Cerro Paranal at a magnitude of 25. The new orbit determination (see 
     Table~\ref{elements}) is based on 70 observations that span a data-arc of 9\,339 d or 25.57 yr and shows that 1991~VG is an Apollo 
     asteroid following a somewhat Earth-like orbit ---semimajor axis, $a$=1.026~au, eccentricity, $e$=0.04975, and inclination, 
     $i$=1\fdg44--- with a minimum orbit intersection distance (MOID) with the Earth of 0.0053~au.
%
%--------------------------------------------------------------------------------------------------------------- Orbital elements of 1991 VG              
%
     \begin{table}
        \fontsize{8}{11pt}\selectfont
        \tabcolsep 0.05truecm
        \caption{Heliocentric Keplerian orbital elements and 1$\sigma$ uncertainties of 1991~VG at epoch JD 2458000.5 that corresponds to 
                 00:00:00.000 TDB, Barycentric Dynamical Time, on 2017 September 4 (J2000.0 ecliptic and equinox. Source: JPL's Small-Body 
                 Database.)
                }
        \centering
        \begin{tabular}{lcc}
           \hline
            Orbital parameter                                 &   & Value$\pm$uncertainty (1$\sigma$) \\ 
           \hline
            Semimajor axis, $a$ (au)                          & = &   1.0255840443$\pm$0.0000000006   \\
            Eccentricity, $e$                                 & = &   0.049746422$\pm$0.000000010     \\
            Inclination, $i$ (\degr)                          & = &   1.437055$\pm$0.000004           \\
            Longitude of the ascending node, $\Omega$ (\degr) & = &  73.26393$\pm$0.00007             \\
            Argument of perihelion, $\omega$ (\degr)          & = &  23.96147$\pm$0.00007             \\
            Mean anomaly, $M$ (\degr)                         & = & 246.83514$\pm$0.00002             \\
            Perihelion, $q$ (au)                              & = &   0.974564908$\pm$0.000000010     \\
            Aphelion, $Q$ (au)                                & = &   1.0766031805$\pm$0.0000000006   \\
            Absolute magnitude, $H$ (mag)                     & = &  28.5                             \\
           \hline
        \end{tabular}
        \label{elements}
     \end{table}
%
%-------------------------------------------------------------------------------------------------------------------------------------------
%

     As for the possible origin of 1991~VG and based only on its orbital elements, Scotti \& Marsden (1991) suggested immediately that it 
     might be a returning spacecraft. West et al. (1991) pointed out that its light curve might be compatible with that of a rapidly 
     rotating satellite ---probably tumbling--- with highly reflective side panels, further supporting the theory that 1991~VG could be an 
     artificial object. Although an artificial origin was proposed first, T.~Gehrels pointed out that if 1991~VG was natural, it might be a 
     representative of a new orbital class of objects, the Arjunas; an unofficial dynamical group of small NEOs following approximately 
     Earth-like orbits which could be secondary fragments of asteroids that were originally part of the main belt and left their formation 
     region under the effect of Jupiter's gravity (Cowen 1993; Rabinowitz et al. 1993; Gladman, Michel \& Froeschl{\'e} 2000). 

     In his analysis of terrestrial impact probabilities, Steel (1995b) assumed that 1991~VG was a returned spacecraft. An artificial nature 
     for 1991~VG was discussed in detail by Steel (1995a), concluding that it was a robust candidate alien artefact (inert or under control); 
     this conclusion was contested by Weiler (1996). The controversy on a possible alien origin for 1991~VG was continued by Steel (1998) 
     and Weiler (1998) to conclude that either the detection of 1991~VG was a statistical fluke (unusual NEO or space debris of terrestrial 
     origin) or a very large number of alien probes are following heliocentric orbits. 

     Tancredi (1997) reviewed all the available evidence to conclude that 1991~VG could be a piece of lunar ejecta, the result of a 
     relatively large impact. Tatum (1997) favoured a natural origin for 1991~VG to state that any asteroid moving in an Earth-like orbit 
     with semimajor axis in the range 0.9943--1.0057~au will inevitably collide with our planet, i.e. observed NEOs in such paths must 
     be relatively recent arrivals. Brasser \& Wiegert (2008) re-examined the topic of the origin of 1991~VG and argued that it had to have 
     its origin on a low-inclination Amor- or Apollo-class object.

     Here and in order to identify the most Earth-like orbits among those of known NEOs, we have used the $D$-criteria of Southworth \& 
     Hawkins (1963), $D_{\rm SH}$, Lindblad \& Southworth (1971), $D_{\rm LS}$ (in the form of equation 1 in Lindblad 1994 or equation 1 in 
     Foglia \& Masi 2004), Drummond (1981), $D_{\rm D}$, and the $D_{\rm R}$ from Valsecchi, Jopek \& Froeschl\'e (1999) to search the known 
     NEOs for objects that could be dynamically similar to our planet, considering the orbital elements of the Earth for the epoch JDTDB 
     2458000.5 (see below) that is the standard time reference used throughout this research. The actual values are: semimajor axis, $a$ = 
     0.999215960~au, eccentricity, $e$ = 0.017237361, inclination, $i$ = 0\fdg000524177, longitude of the ascending node, $\Omega$ = 
     230\fdg950190495, and argument of perihelion, $\omega$ = 233\fdg858793714. The list of NEOs has been retrieved from JPL's Solar System 
     Dynamics Group (SSDG) Small-Body Database (SBDB).\footnote{\url{http://ssd.jpl.nasa.gov/sbdb.cgi}}
 
     Considering the currently available data on NEOs, the orbit of 1991~VG is neither the (overall) most Earth-like known ---the current 
     record holder is 2006~RH$_{120}$ (Bressi et al. 2008a; Kwiatkowski et al. 2009) which has the lowest values of the $D$-criteria, but it 
     has not been observed since 2007 June 22--- nor the one with the orbital period closest to one Earth year ---which is 2014 OL$_{339}$ 
     with $a$=0.9992~au or an orbital period of 364.83227$\pm$0.00002~d (Vaduvescu et al. 2014, 2015; de la Fuente Marcos \& de la Fuente 
     Marcos 2014; Holmes et al. 2015)--- nor the least eccentric ---which is 2002~AA$_{29}$ with $e$=0.01296 (Connors et al. 2002; Smalley 
     et al. 2002), followed by 2003~YN$_{107}$ with $e$=0.01395 (McNaught et al. 2003; Connors et al. 2004)--- nor the one with the 
     lowest inclination ---which is probably 2009~BD with $i$=0\fdg38 (Buzzi et al. 2009; Micheli, Tholen \& Elliott 2012), followed by 
     2013~BS$_{45}$ with $i$=0\fdg77 (Bressi, Scotti \& Hug 2013; de la Fuente Marcos \& de la Fuente Marcos 2013). Other NEOs with very 
     Earth-like orbits are 2006~JY$_{26}$ (McGaha et al. 2006; Brasser \& Wiegert 2008; de la Fuente Marcos \& de la Fuente Marcos 2013) and 
     2008~KT (Gilmore et al. 2008; de la Fuente Marcos \& de la Fuente Marcos 2013). In summary and within the context of the known NEOs, 
     although certainly unusual, the orbital properties of 1991~VG are not as remarkable as initially thought. On a more technical side, the 
     orbit determinations of 2006~RH$_{120}$ and 2009~BD required the inclusion of non-gravitational accelerations (radiation-pressure 
     related) in order to reproduce the available astrometry (see e.g. Micheli et al. 2012). The orbital solution of 1991~VG (see 
     Table~\ref{elements}) was computed without using any non-gravitational forces and reproduces the observations used in the fitting.

  \section{Integrations and orbital evolution}
     Understanding the origin and current dynamical status of 1991~VG demands a detailed study of its past, present and future orbital 
     evolution. A careful statistical analysis of the behaviour over time of its orbital elements and other relevant parameters using a 
     sufficiently large set of $N$-body simulations, including the uncertainties associated with the orbit determination in a consistent 
     manner, should produce reasonably robust conclusions. 

     In this section, we use a publicly available direct $N$-body code\footnote{\url{http://www.ast.cam.ac.uk/~sverre/web/pages/nbody.htm}}
     originally written by Aarseth (2003) that implements a fourth order version of the Hermite scheme described by Makino (1991). The 
     suitability of this software for Solar system studies has been successfully and extensively tested (see de la Fuente Marcos \& de la 
     Fuente Marcos 2012). Consistent with the new data on 1991~VG (Hainaut et al. 2017), non-gravitational forces have been excluded from 
     the calculations; the effects of solar radiation pressure have been found to be negligible in the calculation of the orbit in 
     Table~\ref{elements} and for an average value for the Yarkovsky drift of 10$^{-9}$~au~yr$^{-1}$ (see e.g. Nugent et al. 2012), the 
     time-scale to escape the orbital neighbourhood of the Earth is about 12 Myr, which is some orders of magnitude longer than the time 
     intervals discussed in this research. 

     Our calculations make use of the physical model described by de la Fuente Marcos \& de la Fuente Marcos (2012) and of initial 
     conditions ---positions and velocities in the barycentre of the Solar system for the various bodies involved, including 1991~VG--- 
     provided by JPL's \textsc{horizons}\footnote{\url{https://ssd.jpl.nasa.gov/?horizons}} (Giorgini et al. 1996; Standish 1998; Giorgini \& 
     Yeomans 1999; Giorgini, Chodas \& Yeomans 2001; Giorgini 2011, 2015) at epoch JD 2458000.5 (2017-September-04.0 TDB, Barycentric 
     Dynamical Time), which is the $t$ = 0 instant in our figures. 
%
%-------------------------------------------------------------------------------------------------------------------------------------------
%
     \begin{figure}
        \centering
        \includegraphics[width=\linewidth]{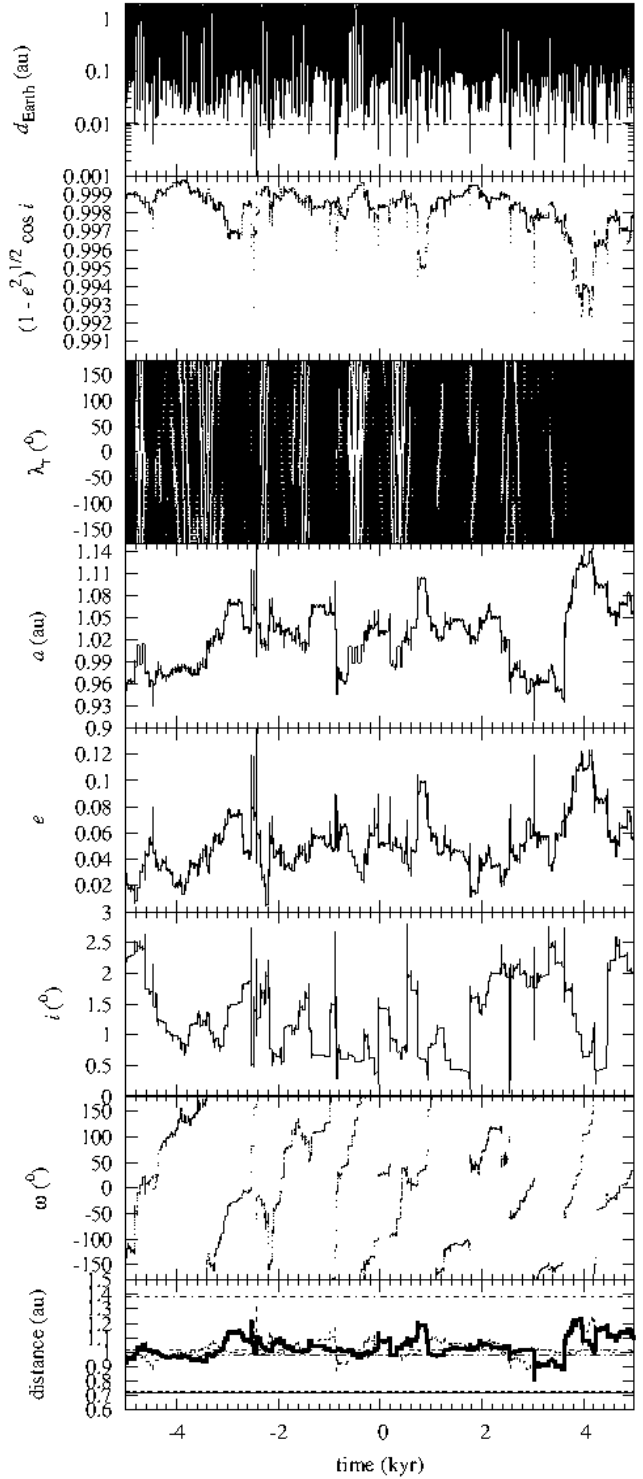}
        \caption{Evolution of the values of the orbital elements and other relevant parameters for the nominal orbit of 1991~VG in 
                 Table~\ref{elements}. The top panel shows the geocentric distance; encounters at ranges well below the Hill radius of 
                 the Earth, 0.0098~au (also shown), are common. The Kozai-Lidov parameter is shown in the second to top panel. The value of
                 the resonant angle is displayed in the third to top panel. The evolution of the orbital elements, semimajor axis, 
                 eccentricity, inclination and argument of perihelion is shown in the fourth to top panel and the fourth, third and second 
                 to bottom panels, respectively. The bottom panel shows the distance from the Sun to the descending (thick line) and 
                 ascending nodes (dotted line); the aphelion and perihelion distances of Venus, the Earth and Mars are indicated as well.
                }
        \label{control}
     \end{figure}
%
%-------------------------------------------------------------------------------------------------------------------------------------------
%
      
     Fig.~\ref{control} shows the evolution backwards and forward in time of several orbital elements and other relevant parameters of 
     1991~VG using initial conditions compatible with the nominal orbit in Table~\ref{elements}. The time interval displayed (10 kyr) is  
     consistent with the analysis of its dynamical lifetime carried out by Brasser \& Wiegert (2008). Fig.~\ref{control}, top panel 
     (geocentric distance), shows that this NEO experiences recurrent close encounters with our planet well within the Hill radius, which is 
     0.0098~au. The Hill radius of the Earth is the maximum orbital distance of an object (natural or artificial) to remain gravitationally 
     bound to our planet, i.e. to be a satellite. When inside the Hill sphere, the Earth's attraction dominates that of the Sun even for 
     objects with a positive value of the geocentric energy, i.e. unbound passing objects. In order to be captured as a satellite of our 
     planet, the geocentric energy of the object must be negative (Carusi \& Valsecchi 1979). This simple criterion does not include any 
     constraint on the duration of the capture event; Rickman \& Malmort (1981) recommended the addition of an additional restriction that 
     the object completes at least one revolution around our planet while its geocentric energy is still negative (see Section 5 for a more 
     detailed analysis applied to 1991~VG).
     
     Given its low eccentricity, 1991~VG cannot undergo close encounters with major bodies other than the Earth--Moon system. As the orbit 
     of 1991~VG is somewhat Earth-like, these are often low-velocity encounters (as low as 0.9~km~s$^{-1}$). Such fly-bys can be very 
     effective in perturbing an orbit even if the close approaches are relatively distant, but in this case we observe very frequent (every 
     few decades) close fly-bys. Under such conditions, one may expect a very chaotic evolution as confirmed by the other panels in 
     Fig.~\ref{control}. Although very chaotic orbits present great challenges in terms of reconstructing the past dynamical evolution of 
     the affected objects or making reliable predictions about their future behaviour, Wiegert, Innanen \& Mikkola (1998) have shown that it 
     is still possible to arrive to scientifically robust conclusions if a proper analysis is performed. On the other hand, low-velocity 
     encounters well within the Hill radius can lead to temporary capture events (see Section 5). 

\begin{figure}
        \centering
        \includegraphics[width=\linewidth]{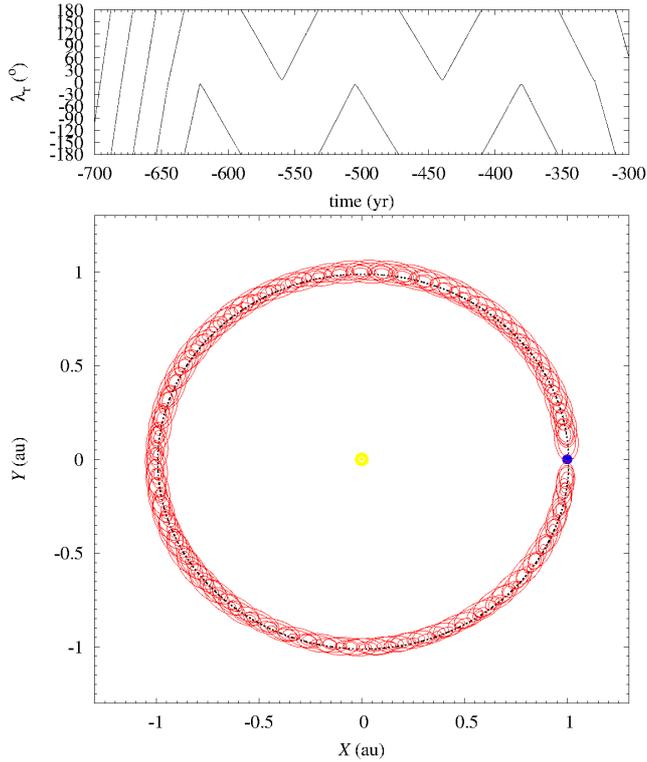}
        \caption{Variation of the relative mean longitude, $\lambda_{\rm r}$, over time during the most recent co-orbital episode of 1991~VG 
                 (top panel). The path followed by 1991~VG during the time interval ($-$635, $-$350)~yr (bottom panel) describes a horseshoe 
                 pattern when seen in a frame of reference centred at the Sun and rotating with the Earth, projected on to the ecliptic 
                 plane. The figure also shows the orbit of the Earth, its position at (1, 0)~au, and the Sun at (0, 0)~au.  
                }
        \label{orbit}
     \end{figure}

     Fig.~\ref{control}, second to top panel, shows the evolution of the value of the so-called Kozai-Lidov parameter $\sqrt{1 - e^2} \cos i$ 
     (Kozai 1962; Lidov 1962) that measures the behaviour of the component of the orbital angular momentum of the minor body perpendicular 
     to the ecliptic. The value of this parameter remains fairly constant over the time interval studied; the dispersion is smaller than the 
     one observed for typical NEOs following Earth-like paths (see e.g. figs 3, 6 and 9, B-panels, in de la Fuente Marcos \& de la Fuente 
     Marcos 2016b). The variation of the relative mean longitude of 1991~VG or difference between the mean longitude of the object and that 
     of the Earth, $\lambda_{\rm r}$ (see e.g. Murray \& Dermott 1999), is shown in Fig.~\ref{control}, third to top panel. When 
     $\lambda_{\rm r}$ changes freely in the interval (0, 360)\degr ---i.e. $\lambda_{\rm r}$ circulates--- 1991~VG is not subjected to the 
     1:1 mean-motion resonance with our planet. If the value of $\lambda_{\rm r}$ oscillates or librates ---about 0{\degr} (quasi-satellite), 
     $\pm$60{\degr} (Trojan) or 180{\degr} (horseshoe)--- then the orbital periods of 1991~VG and the Earth are virtually the same and we 
     speak of a co-orbital companion to our planet. Fig.~\ref{control}, third to top panel, shows that 1991~VG has been a recurrent transient 
     co-orbital of the horseshoe type in the past and it will return as such in the future, however, it is not a present-day co-orbital 
     companion of the Earth. Fig.~\ref{orbit} shows the most recent co-orbital episode of 1991~VG in further detail; the variation of the 
     relative mean longitude indicates that 1991~VG followed a horseshoe path for about 300~yr. 
%
%-------------------------------------------------------------------------------------------------------------------------------------------
%
%
%-------------------------------------------------------------------------------------------------------------------------------------------
%

     Fig.~\ref{control}, fourth to top panel, shows the evolution of the value of the semimajor axis of 1991~VG. Earth's co-orbital region
     goes from $\sim$0.994~au to $\sim$1.006~au, or a range in orbital periods of 362--368~d, and the figure shows that 1991~VG only enters 
     this zone for relatively brief periods of time although it remains in its neighbourhood during the entire integration. 
     Fig.~\ref{control}, fourth and third to bottom panels, shows how the eccentricity and the inclination, respectively, change over time. 
     Although in both cases the evolution is very irregular, there is some weak coupling between both orbital elements and in some cases, 
     when the eccentricity reaches a local maximum, the value of the inclination reaches a local minimum and vice versa. This explains why 
     the value of the Kozai-Lidov parameter (Fig.~\ref{control}, second to top panel) remains relatively stable throughout the integrations; 
     this is also a sign that the Kozai-Lidov mechanism (Kozai 1962; Lidov 1962) may be at work, at least partially. This interpretation is 
     confirmed in Fig.~\ref{control}, second to bottom panel, as the value of the argument of perihelion, $\omega$, shows signs of libration 
     (it does not circulate) which is a typical side effect of the Kozai-Lidov mechanism. Fig.~\ref{control}, bottom panel, shows the 
     evolution of the nodal distances of 1991~VG; encounters with the Earth--Moon system are only possible in the neighbourhood of the nodes 
     and both nodes tend to drift into the path of our planet in a chaotic manner.  
      
\begin{figure}
        \centering
        \includegraphics[width=\linewidth]{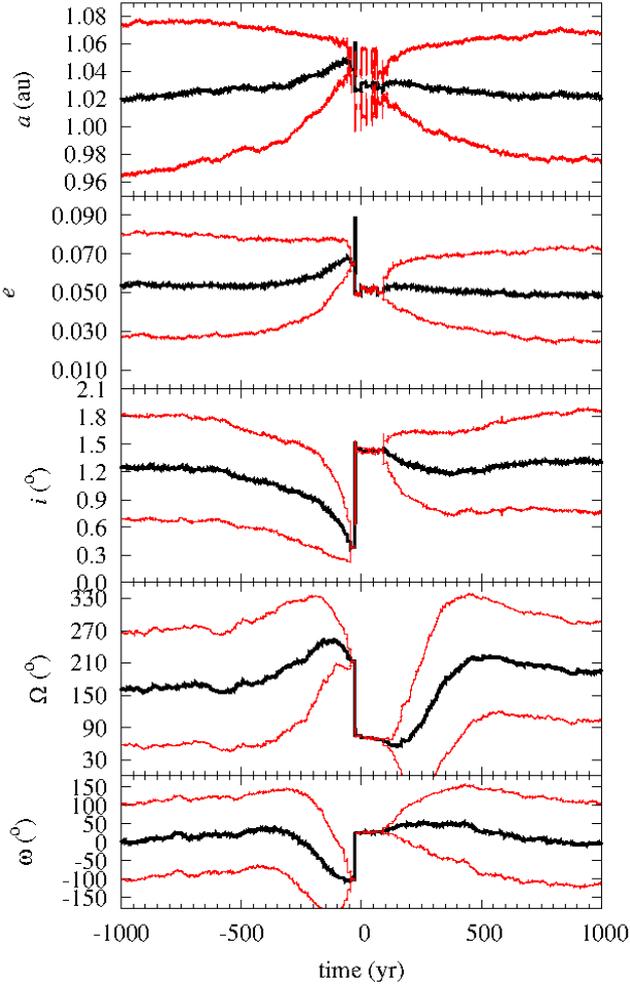}
        \caption{Evolution of the dispersions of the values of the orbital elements of 1991~VG for 250 control orbits: semimajor axis (top 
                 panel), eccentricity (second to top panel), inclination (middle panel), longitude of the ascending node (second to bottom 
                 panel), and argument of perihelion (bottom panel). Average values are plotted as thick black curves and their ranges 
                 (1$\sigma$ uncertainties) as thin red curves.  
                }
        \label{disper}
     \end{figure}

     The orbital evolution displayed in Fig.~\ref{control} gives a general idea of the dynamical behaviour of 1991~VG, but it does not show
     the effect of the uncertainties in the orbit determination (see Table~\ref{elements}). In order to account for this critical piece of 
     information, we use the Monte Carlo using the Covariance Matrix (MCCM) method detailed in section 3 of de la Fuente Marcos \& de la 
     Fuente Marcos (2015c) ---the covariance matrix to generate initial positions and velocities has been obtained from JPL's 
     \textsc{horizons}. Fig.~\ref{disper} shows the results of the evolution of 250 control orbits generated using the MCCM method. These
     simulations confirm that the orbital evolution of 1991~VG is chaotic on time-scales longer than a few decades.
%
%-------------------------------------------------------------------------------------------------------------------------------------------
%
%
%-------------------------------------------------------------------------------------------------------------------------------------------
%

  \section{Unusual but not uncommon}
     One of the arguments originally applied to reject a natural origin for 1991~VG was that, in accordance with the evidence available at 
     that time, such a NEO was highly improbable, from an orbital point of view. Brasser \& Wiegert (2008) using numerical simulations 
     estimated that the probability of a NEO ever ending up on an Earth-like orbit could be about 1:20\,000. If we use the latest NEO 
     orbital model described by Granvik et al. (2013a,b) and Bottke et al. (2014) and implemented in the form of a publicly available survey 
     simulator,\footnote{\url{http://neo.ssa.esa.int/neo-population}} we obtain a probability of finding a NEO moving in an orbit akin to 
     that of 1991~VG of about $10^{-6}$ ---the degree of similarity between two orbits has been estimated as described before, using the 
     $D$-criteria. As the previously mentioned NEO orbit model only strictly applies to NEOs with $H<25$~mag (in fact, the single object 
     predicted by the model has a magnitude of 24.24) and 1991~VG is smaller, it cannot be discarded that objects other than 1991~VG may be 
     moving along similar paths. 

     In order to confirm or reject this plausible hypothesis, we have used the $D$-criteria mentioned above to search the known NEOs (data 
     from JPL's SSDG SBDB as before) for objects dynamically similar to 1991~VG. We apply these criteria using osculating Keplerian orbital 
     elements, not the customary proper orbital elements (see e.g. Milani 1993, 1995; Milani \& Kne{\v z}evi{\'c} 1994; Kne{\v z}evi{\'c} 
     \& Milani 2000; Milani et al. 2014, 2017), because 1991~VG-like orbits are inherently chaotic on very short time-scales. 
     Table~\ref{aliens} shows the data of three other NEOs ---2001~GP$_{2}$ (McMillan et al. 2001), 2008~UA$_{202}$ (Bressi et al. 2008b) 
     and 2014~WA$_{366}$ (Gibbs et al. 2014)--- which are dynamically similar to 1991~VG (they have $D_{\rm LS}$ and $D_{\rm R} < 0.05$). 
     Integrations analogous to those in Fig.~\ref{control} (not shown) indicate that the evolution of these three NEOs (although their orbit 
     determinations are in need of significant improvement) bears some resemblance to that of 1991~VG. Common features include relatively 
     frequent close encounters with the Earth--Moon system and very chaotic short-term evolution (all of them), small values of their MOIDs, 
     recurrent trapping by 1:1 mean-motion resonances with the Earth (particularly 2008~UA$_{202}$), evolution temporarily affected by the 
     Kozai-Lidov effect and other consistent properties. Asteroid 2008~UA$_{202}$ is considered an easily retrievable NEO (Garc{\'{\i}}a 
     Y{\'a}rnoz, Sanchez \& McInnes 2013).  
%
%-------------------------------------------------------------------------------------------------------------------------------- Candidates
%
     \begin{table*}
        \centering
        \fontsize{8}{11pt}\selectfont
        \tabcolsep 0.07truecm
        \caption{Orbital elements, orbital periods ($P$), perihelion ---$q = a \ (1 - e)$--- and aphelion ---$Q = a \ (1 + e)$--- distances, 
                 number of observations ($n$), data-arc span, absolute magnitudes ($H$) and MOID with the Earth of NEOs following orbits 
                 similar to that of 1991~VG. The values of the various $D$-criteria ($D_{\rm SH}$, $D_{\rm LS}$, $D_{\rm D}$ and 
                 $D_{\rm R}$) with respect to 1991~VG are displayed as well. The minor bodies are sorted by ascending $D_{\rm LS}$ and only 
                 those with $D_{\rm LS}$ and $D_{\rm R} < 0.05$ are shown. The orbits are referred to the epoch 2017 September 4 as before. 
                 Source: JPL's Small-Body Database.}
        \begin{tabular}{lllllllllllllllll}
           \hline
            Asteroid        & $a$ (au)  & $e$        & $i$ (\degr) & $\Omega$ (\degr) & $\omega$ (\degr) & $P$ (yr) & $q$ (au) & $Q$ (au)
                            & $n$ & arc (d) & $H$ (mag) & MOID (au)
                            & $D_{\rm SH}$ & $D_{\rm LS}$ & $D_{\rm D}$ & $D_{\rm R}$ \\
           \hline
            2014 WA$_{366}$ & 1.03433   & 0.07150    & 1.55915     &  67.10073        & 287.63348        &  1.05    & 0.9604   & 1.1083 
                            & 55  & 49      & 26.9      & 0.00732
                            & 0.0981       & 0.0261       & 0.1829      & 0.0276      \\
            2001 GP$_{2}$   & 1.03779   & 0.07380    & 1.27825     & 196.80669        & 111.40484        &  1.06    & 0.9612   & 1.1144 
                            & 25  &  27     & 26.9      & 0.00174
                            & 0.1291       & 0.0277       & 0.2019      & 0.0201      \\
            2008 UA$_{202}$ & 1.03318   & 0.06857    & 0.26357     &  21.08289        & 300.90518        &  1.05    & 0.9623   & 1.1040 
                            & 16  &  6      & 29.4      & 0.00022
                            & 0.1139       & 0.0304       & 0.1655      & 0.0132      \\
           \hline
        \end{tabular}
        \label{aliens}
     \end{table*}
%
%-------------------------------------------------------------------------------------------------------------------------------------------
%

     NEOs 2001 GP$_{2}$ and 2008 UA$_{202}$ are included in the list of asteroids that may be involved in potential future Earth impact 
     events maintained by JPL's Sentry System (Chamberlin et al. 2001; Chodas 2015).\footnote{\url{https://cneos.jpl.nasa.gov/sentry/}} 
     Asteroid 2001 GP$_{2}$ has a computed impact probability of 0.00021 for a possible impact in 2043--2107; asteroid 2008 UA$_{202}$ is
     listed with an impact probability of 0.000081 for a possible impact in 2050--2108. Their very similar range of years for a most 
     probable potential impact ---i.e. they reach their closest perigees at similar times even if their synodic periods are long--- suggest 
     a high degree of dynamical coherence for these two objects even if their values of $\Omega$ and $\omega$ are very different. Although 
     they have relatively high values of the impact probability, their estimated diameters are smaller than 20~m. In the improbable event of 
     an impact, its effects would be local, not too different from those of the Chelyabinsk event or some other recent minor impacts (see 
     e.g. Brown et al. 2013; Popova et al. 2013; de la Fuente Marcos \& de la Fuente Marcos 2015b; de la Fuente Marcos, de la Fuente Marcos 
     \& Mialle 2016); however, an object like 2008 UA$_{202}$ probably would break up in the Earth's atmosphere and few fragments, if any, 
     will hit the ground (or the ocean). 

  \section{Natural or artificial?}
     Some of the orbital properties of 1991~VG have been used to argue in favour or against a natural or artificial origin for this object.
     A very unusual dynamical feature that was pointed out by Tancredi (1997) was the fact that 1991~VG experienced a temporary satellite 
     capture by the Earth during its 1991--1992 fly-by; such satellite capture showed a recurrent pattern. 

\begin{figure}
        \centering
        \includegraphics[width=\linewidth]{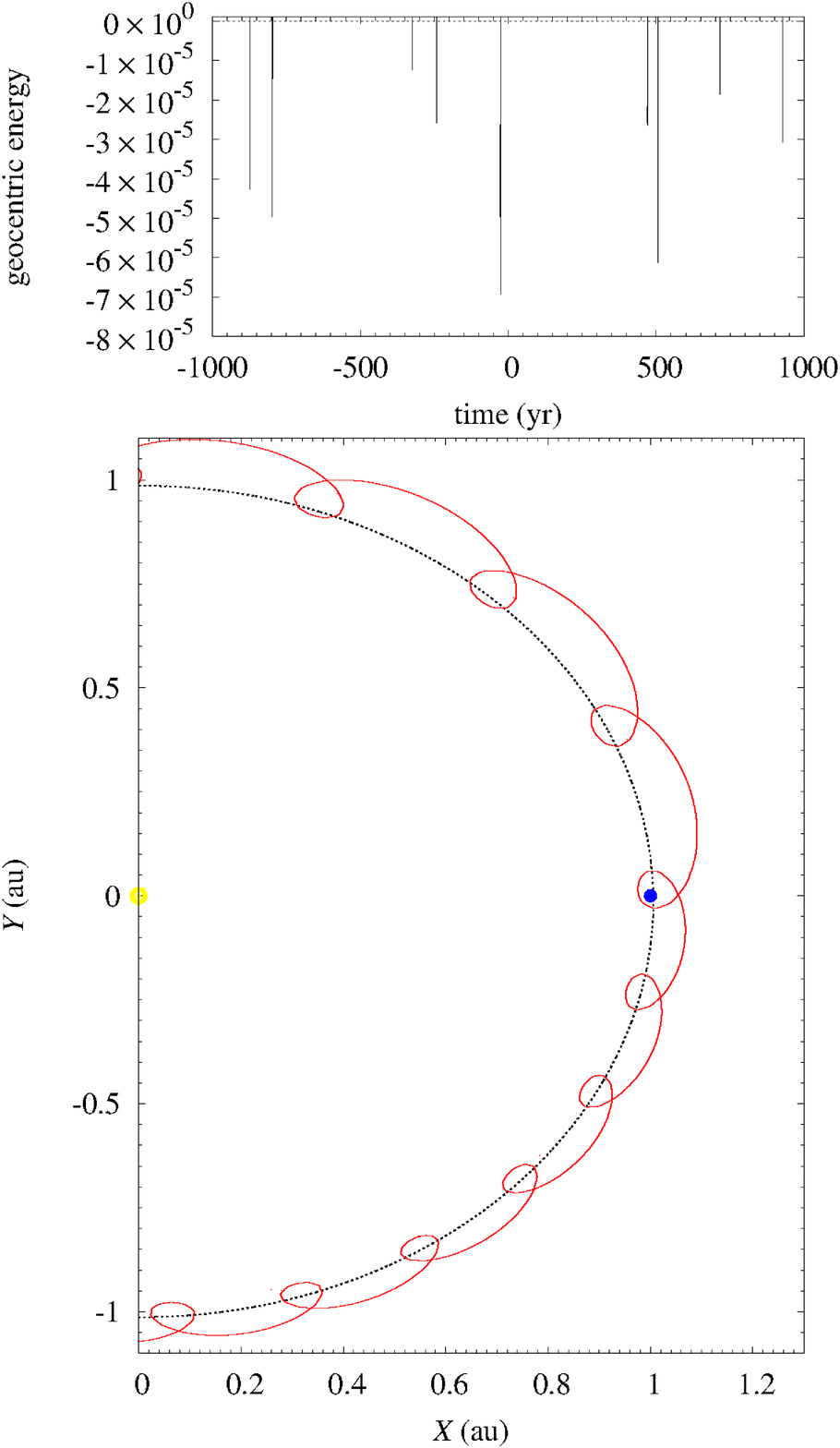}
        \caption{Keplerian geocentric energy of 1991~VG as a function of time (top panel), ephemeral (lasting about a month) satellite 
                 captures happen when the value of the geocentric energy becomes negative. The unit of energy is such that the unit of mass 
                 is 1~$M_{\odot}$, the unit of distance is 1~au and the unit of time is one sidereal year divided by 2$\pi$. The path 
                 followed by 1991~VG (same frame of reference as in Fig.~\ref{orbit}) during the approximate time interval ($-$30, $-$19)~yr 
                 (bottom panel) shows that this minor body went around our planet once during its previous fly-by in 1991--1992. Asteroid 
                 1991~VG moves clockwise (retrograde) in the figure as the time goes forward; the temporary satellite capture event happened 
                 during 1992 February. 
                }
        \label{energy}
     \end{figure}

     The backward evolution of the new orbit determination fully confirms the analysis made by Tancredi (1997). Fig.~\ref{energy}, top 
     panel, shows that the Keplerian geocentric energy of 1991~VG (relative binding energy) became negative during the encounter and also 
     that 1991~VG completed an entire revolution around our planet (bottom panel). Therefore, it might have matched both criteria (see 
     above, Carusi \& Valsecchi 1979; Rickman \& Malmort 1981) to be considered a bona fide satellite or, perhaps more properly, a minimoon 
     of the Earth; using the terminology in Fedorets, Granvik \& Jedicke (2017) we may speak of a temporarily captured orbiter. However, the 
     relative binding energy was not negative for the full length of the loop around our planet pictured in Fig.~\ref{energy}, bottom panel.
     As the loop followed by 1991~VG with respect to the Earth was travelled in the clockwise sense, this event may be regarded as the first 
     ever documented retrograde capture of a satellite by our planet, even if it had a duration of about 28 d. Fig.~\ref{energy}, top panel, 
     also shows that this unusual phenomenon is recurrent in the case of 1991~VG. But, if there are other objects moving in 1991~VG-like 
     orbits, how often do they become temporary satellites of the Earth?
%
%-------------------------------------------------------------------------------------------------------------------------------------------
%
%
%-------------------------------------------------------------------------------------------------------------------------------------------
%

\begin{figure}
        \centering
         \includegraphics[width=\linewidth]{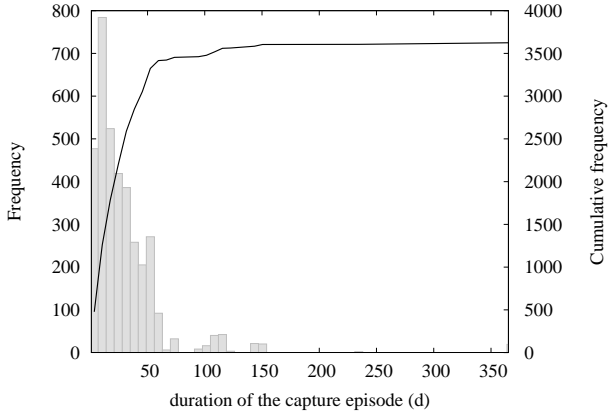}
         \caption{Frequency distribution of the duration of episodes of temporary capture as natural satellite of our planet. The bin size 
                  is 7 d.
                 }
         \label{dminimoons}
      \end{figure}

\begin{figure}
        \centering
         \includegraphics[width=\linewidth]{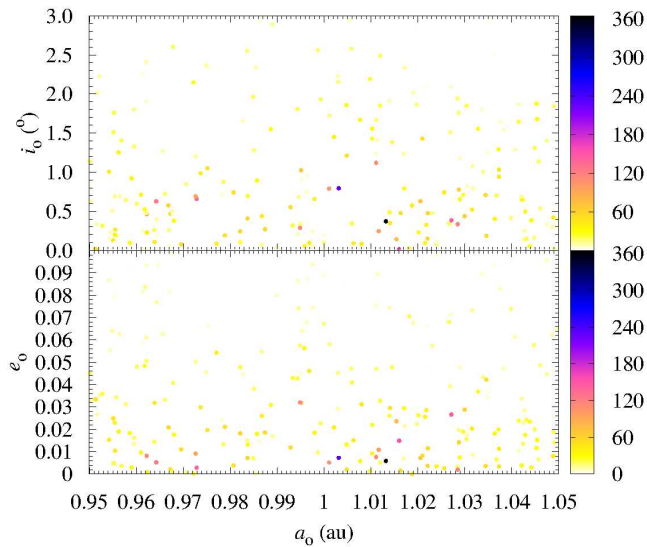}
         \caption{Duration of episodes of temporary capture as natural satellite of our planet in days as a function of the initial values 
                  of $a$ and $e$ (bottom panel) and $a$ and $i$ (top panel). The colours (or grey scale) in the colour maps are proportional 
                  to the duration in days of the episode in Fig.~\ref{dminimoons}. The results of $10^{6}$ experiments are plotted (see the 
                  text for details).
                 }
         \label{mapminimoons}
      \end{figure}

     Figs~\ref{dminimoons} and \ref{mapminimoons} show the results of $10^{6}$ numerical experiments in which a virtual object moving in 
     a 1991~VG-like orbit ---orbital elements assumed to be uniformly distributed in the volume of orbital parameter space defined by 
     $a\in(0.95, 1.05)$~au, $e\in(0.0, 0.1)$, $i\in(0, 3)$\degr, $\Omega\in(0, 360)$\degr, $\omega\in(0, 360)$\degr, and the time of 
     perihelion passage $\tau_{q}\in(2458000.5, 2458365.75)$ JD--- undergoes a fly-by with our planet. The region chosen encloses the orbit 
     solutions of 1991~VG, 2001~GP$_{2}$, 2008~UA$_{202}$ and 2014~WA$_{366}$, and those of other NEOs cited earlier in this paper as well. 
     We did not use the NEO orbit model mentioned before to generate the synthetic orbits because we wanted to survey the relevant volume of 
     orbital parameter space in full detail so our results could be applied to both natural and artificial objects. The evolution of the 
     virtual objects was followed just for one year of simulated time to minimize the impact of orbital chaos and resonant returns (see e.g. 
     Milani, Chesley \& Valsecchi 1999) on our conclusions. This short time interval is fully justified because our previous analyses show 
     that, after experiencing a close fly-by with the Earth, an object moving in a 1991~VG-like orbit most likely jumps into another 
     1991~VG-like orbit. These experiments have been carried out with the same software, physical model and relevant initial conditions used 
     in our previous integrations. 

     Our calculations show that the probability of becoming a temporary (for any length of time) satellite of our planet for members of this 
     group of objects is 0.0036. The overall capture rate is roughly similar to that found by Granvik et al. (2012). In our case, the 
     probability of a capture for less than one month is 0.0023, for one to two months is 0.0011, and for more than two months is 0.00021 
     (see Fig.~\ref{dminimoons}). Captures for less than 7 d are less probable than captures for 7 to 14 d. Therefore, most objects 
     temporarily captured as Earth's transient bound companions spend less than one month in this state and they do not complete one full 
     revolution around our planet. These results also indicate that as long as we have NEOs in 1991~VG-like orbits, they naturally tend to 
     become temporary satellites of our planet. However, captures for more than two months are rather unusual and those lasting one year, 
     exceedingly rare. This result is at odds with those in Granvik et al. (2012) and Fedorets et al. (2017), but this is not surprising
     because our short integrations are biased against producing temporarily captured orbiters due to the comparatively small size of our 
     synthetic sample ---$10^{6}$ experiments versus $10^{10}$ in Granvik et al. (2012)--- and our choice of initial conditions ---e.g. 
     their integrations start at 4--5 Hill's radii from the Earth. Fedorets et al. (2017) show explicitly in their table 1 that 40 per cent 
     of all captures should be temporarily captured orbiters.  

     As objects moving in 1991~VG-like paths are being kicked from one of these orbits into another, it is perfectly normal to observe 
     recurrent but brief capture episodes as those discussed for 1991~VG. Such temporary captures are also observed during the integrations 
     of 2001~GP$_{2}$, 2008~UA$_{202}$ and 2014~WA$_{366}$ although they tend to be shorter in duration and less frequent. In addition, no 
     virtual object collided with the Earth during the calculations, which strongly suggests that even if impact probabilities are 
     theoretically high for many of them, it is much more probable to be captured as an ephemeral satellite of our planet than to become an 
     actual impactor; this conclusion is consistent with recent results by Clark et al. (2016), but it is again at odds with results 
     obtained by Granvik et al. (2012) and Fedorets et al. (2017), who found that about one per cent of their test particles impacted the 
     Earth. This significant discrepancy comes out of the facts pointed out before, comparatively small size of our synthetic sample and 
     different initial conditions. 
%
%-------------------------------------------------------------------------------------------------------------------------------------------
%
%
%-------------------------------------------------------------------------------------------------------------------------------------------
%

     Fig.~\ref{mapminimoons} shows how the duration of the episode of temporary capture as natural satellite of our planet depends on the 
     initial values of the semimajor axis, eccentricity and inclination. The colours (or grey scale) in the maps depend on the duration of 
     the episode in days as indicated in the associated colour box. NEOs moving in 1991~VG-like orbits of the Amor- or Apollo-class are more 
     likely to experience longer capture episodes. The probability of getting captured decreases rapidly for objects with $e>0.05$ and/or 
     $i>1\fdg5$, and the duration of the recorded episodes is shorter. It is important to notice that one of these virtual objects might 
     leave the assumed initial volume of NEO orbital parameter space (see above) in a time-scale of the order of 1 kyr as Fig.~\ref{disper} 
     shows. In addition, the longest orbital period of a satellite of our planet is about 205~d (if it is at the Hill radius); therefore,
     most of the temporary captures recorded in our numerical experiment do not qualify as true satellites according to Rickman \& Malmort 
     (1981) because they did not complete at least one revolution when bound to the Earth; following the terminology used by Fedorets et al. 
     (2017), we may speak of temporarily captured fly-bys in these cases. In fact and strictly speaking, the event in Fig.~\ref{energy} is
     compatible with a temporarily captured fly-by not a temporarily captured orbiter; one of the annual epicycles happens to (somewhat 
     accidentally) loop around the Earth lasting several months, but the geocentric energy is negative only for a fraction of the time taken
     to travel the loop. Indeed, our limited numerical experiment has been optimized to show how frequent temporarily captured fly-bys 
     ---not orbiters--- are; the histogram in Fig.~\ref{mapminimoons} matches reasonably well that of temporarily captured fly-bys in fig.~2 
     in Fedorets et al. (2017).
%
%-------------------------------------------------------------------------------------------------------------------------------------------
%
%
%-------------------------------------------------------------------------------------------------------------------------------------------
%

     The topic of the capture of irregular satellites by planets has been studied by e.g. Astakhov et al. (2003), Nesvorn{\'y}, 
     Vokrouhlick{\'y} \& Morbidelli (2007) and Emel'yanenko (2015). Jupiter is a well-documented host for these captures (see e.g. Rickman 
     \& Malmort 1981; Tancredi, Lindgren \& Rickman 1990; Kary \& Dones 1996). In regards to the Earth, the topic has only recently received 
     attention (see e.g. Baoyin, Chen \& Li 2010; Granvik, Vaubaillon \& Jedicke 2012; Bolin et al. 2014; Brelsford et al. 2016; Clark et 
     al. 2016; Jedicke et al. 2016; Fedorets et al. 2017). Fedorets et al. (2017) have predicted that the largest body constantly present on 
     a geocentric orbit could have a diameter of the order of 0.8~m. 

     The recent scientific interest on the subject of transient bound companions of our planet was triggered by the exceptional close 
     encounter between our planet and 2006~RH$_{120}$ (Bressi et al. 2008a; Kwiatkowski et al. 2009). Kwiatkowski et al. (2009) showed that 
     2006~RH$_{120}$ was temporarily captured into a geocentric orbit from 2006 July to 2007 July. In their work, they confirmed that 
     2006~RH$_{120}$ is a natural object and it cannot be lunar ejecta; they favour a scenario in which its capture as transient satellite 
     was the result of aerobraking in the Earth's atmosphere of a NEO previously moving in a standard Earth-crossing orbit with very low 
     MOID, a low-eccentricity Amor- or Apollo-class minor body. 

     If we interpret the capture of 2006~RH$_{120}$ as a minimoon within the context of our previous numerical experiment, in which the 
     probability of remaining captured for an entire year is about $10^{-6}$, this episode might be a statistical fluke or perhaps indicate 
     that the population of NEOs capable of experiencing such episodes is exceedingly large. In principle, our results strongly favour the 
     interpretation of the 2006~RH$_{120}$ capture episode as a clear outlier. Short-lived satellite capture events consistent with those in 
     Figs~\ref{dminimoons} and \ref{mapminimoons} have been routinely observed in simulations of real NEOs moving in Earth-like orbits (see 
     e.g. the discussion in de la Fuente Marcos \& de la Fuente Marcos 2013, 2014, 2015a). While our simulations show that the capture 
     episode experienced by 1991~VG is unusual but not uncommon, the one by 2006~RH$_{120}$ seems to be truly uncommon and it is difficult 
     to assume that the same scenario that led 1991~VG to become a minimoon can be applied to 2006~RH$_{120}$ as well. However, as 
     2006~RH$_{120}$ is a confirmed (by radar) natural object, it is reasonable to assume that 1991~VG is natural too. Within the context
     of numerical experiments optimized to study temporarily captured orbiters ---not fly-bys--- the case of 2006~RH$_{120}$ is not unusual 
     though. In fact, Granvik et al. (2012) and Fedorets et al. (2017) found a good match between the probability associated with the 
     capture of 2006~RH$_{120}$ and predictions from their models for the temporarily captured orbiter population.  

     The orbital solution in Table~\ref{elements} predicts that in addition to its close approaches to our planet in 2017--2018 (2017 August 
     7 and 2018 February 11) and 1991--1992 (1991 December 5 and 1992 April 9), 1991~VG experienced similar fly-bys on 1938 August 31 and 
     1939 March 14, then on 1956 June 14 and 1957 March 26, and more recently on 1974 August 27 and 1975 March 15. The first two dates 
     predate the start of any human space exploration programme, so they can be customarily discarded. The date in 1956 follows some 
     suborbital tests, one of them on June 13; the same can be said about 1957, another suborbital flight took place on March 25. It is 
     highly unlikely that debris from suborbital tests could have been able to escape into a heliocentric orbit to return at a later time. 
     There were no documented launches in or around 1975 March 15, but the spacecraft Soyuz 15 was launched on 1974 August 26 at 19:58:05 
     UTC (Clark 1988; Newkirk 1990). This manned spacecraft failed to dock with the Salyut 3 space station due to some electronic 
     malfunction and returned to the ground safely two days later. One may argue that 1991~VG might be some part of Soyuz 15 (perhaps some 
     stage of the Proton K/D launch system) as the time match is very good, but Soyuz 15 followed a low-Earth orbit and it is extremely 
     unlikely that any of the stages of the heavy-lift launch vehicle (e.g. the second stage, 8S11K, of 14 m and 11\,715 kg or the third 
     stage of 6.5 m and 4\,185 kg, empty weights) could have escaped the gravitational field of our planet.

     Although both space debris and active spacecraft have received temporary designations as minor bodies by mistake (see e.g. section 9 of 
     de la Fuente Marcos \& de la Fuente Marcos 2015d), objects with initial conditions coming from artificial paths tend to be removed from 
     Earth's orbital neighbourhood rather fast (de la Fuente Marcos \& de la Fuente Marcos 2015d). This is to be expected as spacecraft 
     move under mission control commands and trajectories must be corrected periodically. In addition, the orbital solutions of the NEOs 
     mentioned in the previous sections (including that of 1991~VG) did not require the inclusion of any non-gravitational acceleration 
     (e.g. pressure related) to reproduce the available observations with the exception of two objects, 2006~RH$_{120}$ and 2009~BD, for 
     which the effects of the radiation pressure were detected (see e.g. Micheli et al. 2012). 

     Objects of artificial origin are characterized by a low value of their bulk density (while the average asteroid density is 
     2\,600~kg~m$^{-3}$, but 2006~RH$_{120}$ has 400~kg~m$^{-3}$ and 2009~BD may have 640~kg~m$^{-3}$, Micheli et al. 2012) or conversely by 
     a high value of its proxy, the Area to Mass Ratio (AMR), that may be $>10^{-3}$~m$^{2}$~kg$^{-1}$ for an artificial object. The lowest 
     values of the bulk density of natural objects are linked to the presence of highly porous rocky materials. The density of a fully 
     loaded 8S11K was about 886 kg m$^{-3}$ and an empty one was significantly less dense at perhaps 62 kg m$^{-3}$ (as a hollow metallic 
     shell); the AMR of an empty 8S11K is close to 5.4$\times$10$^{-3}$~m$^{2}$~kg$^{-1}$. The AMR values of the NEOs cited in this work 
     are compatible with a natural origin for all of them (including 1991~VG). Reproducing the paths followed by objects of artificial 
     origin (e.g. space debris) requires the inclusion of non-gravitational accelerations in order to properly account for the observational 
     data; this is also applicable to inert or active spacecraft and very likely to any putative extraterrestrial artefact. As 1991~VG does
     not exhibit any of the properties characteristic of artificial objects, it must be natural.

  \section{Discussion}
     The data review and analyses carried out in the previous sections show that, although certainly unusual, the orbital properties and the 
     dynamical evolution of 1991~VG are not as remarkable as originally thought. Although initially regarded as a mystery object, it is in 
     fact less of a puzzle and more of a dynamically complex NEO, one of a few dozens which roam temporarily in the neighbourhood of the 
     path of the Earth.

     Steel (1995a, 1998) hypothesized that 1991~VG could be an alien-made object, some type of self-replicating probe akin to that in von 
     Neumann's concept. In perspective, this imaginative conjecture may indeed be compatible with what is observed in the sense that if an 
     inert fleet of this type of alien-made objects is actually moving in Earth-like orbits, they would behave as an equivalent population 
     of natural objects (i.e. NEOs) moving in 1991~VG-like orbits. The presence of a present-day active (i.e. under intelligent control) 
     fleet of alien probes can be readily discarded because the observed objects do not appear to be subjected to any non-gravitational 
     accelerations other than those linked to radiation pressure and perhaps the Yarkovsky effect, and also because of the lack of detection 
     of any kind of alien transmissions. An inert (or in hibernation mode) fleet of extraterrestrial artefacts would be dynamically 
     indistinguishable from a population of NEOs if the values of their AMRs are low enough. However and adopting Occam's razor, when there 
     exist two explanations for an observed event, the simpler one must be given precedence. Human space probes, radar, spectroscopic and 
     photometric observations performed over several decades have all shown that, in the orbital neighbourhood of the Earth, it is far more 
     common to detect space rocks than alien artefacts. 

     Scotti \& Marsden (1991) and West et al. (1991) used orbital and photometric data to argue that 1991~VG could be a human-made object,
     a piece of rocket hardware or an old spacecraft that was launched many decades ago. The orbital evolutions of relics of human space 
     exploration exhibit a number of traits that are distinctive, if not actually unique, and the available observational evidence indicates
     that none of them are present in the case of 1991~VG and related objects; almost certainly, 1991~VG was never launched from the Earth. 
     The putative fast rotation period and large-amplitude light curve reported in West et al. (1991) could be compatible with 1991~VG being 
     the result of a relatively recent fragmentation event, where the surface is still fresh, and an elongated boulder is tumbling rapidly; 
     2014 WA$_{366}$ has an orbit quite similar to that of 1991~VG, perhaps they are both fragments of a larger object.

     Tancredi (1997) put forward a novel hypothesis for the origin of 1991~VG, ejecta from a recent lunar impact. Our calculations strongly
     suggest that objects moving in 1991~VG-like orbits may not be able to remain in this type of orbit for an extended period of time. 
     These integrations indicate that, perhaps, the present-day 1991~VG is less than 10 kyr old in dynamical terms, but impacts on the Moon
     capable of ejecting objects the size of 1991~VG are nowadays very, very rare. Brasser \& Wiegert (2008) have studied this issue in 
     detail and the last time that a cratering event powerful enough to produce debris consistent with 1991~VG took place on the Moon could 
     be about 1 Myr ago. Unless we assume that 1991~VG was born that way, then left the orbital neighbourhood of the Earth, and recently was 
     reinserted there, the presence of 1991~VG is difficult to reconcile with an origin as lunar ejecta.

     After discarding an artificial (alien or human) or lunar origin, the option that remains is the most natural one, 1991~VG could be an 
     unusual but not uncommon NEO. We know of dozens of relatively well-studied NEOs that move in Earth-like orbits and our analysis shows 
     that three of them follow 1991~VG-like orbits. All these orbits are characterized by relatively high probabilities of experiencing 
     temporary captures as satellites of the Earth and also of becoming trapped in a 1:1 mean motion resonance with our planet; in a 
     recurrent manner for both cases. We have robust numerical evidence that 1991~VG has been a co-orbital and a satellite of our planet in 
     the past and our calculations predict that these events will repeat in the future. Therefore, the peculiar dynamics of 1991~VG is not 
     so remarkable after all, when studied within the context of other NEOs moving in Earth-like orbits. In addition to the few discussed 
     here, multiple examples of these behaviours can be found in the works by de la Fuente Marcos \& de la Fuente Marcos (2013, 2015a,b, 
     2016a,b). NEOs moving in 1991~VG-like orbits tend to spend less than one month as satellites (i.e. inside the Hill sphere) of our 
     planet and follow paths of the horseshoe type when moving co-orbital (i.e. outside the Hill radius) to our planet; they appear to avoid 
     the Trojan and quasi-satellite resonant states (see e.g. de la Fuente Marcos \& de la Fuente Marcos 2014, 2015a) perhaps because of 
     their comparatively wide semimajor axes relative to that of the Earth. Another unusual dynamical property of 1991~VG and related 
     objects is that of being subjected to the Kozai-Lidov mechanism (see Fig.~\ref{control}, second to bottom panel, libration in 
     $\omega$), at least for brief periods of time. This is also a common behaviour observed for many NEOs moving in Earth-like orbits (see 
     e.g. de la Fuente Marcos \& de la Fuente Marcos 2015d).    
 
     In a seminal work, Rabinowitz et al. (1993) argued that 1991~VG and other NEOs were signalling the presence of a secondary asteroid 
     belt around the path of our planet. These objects are unofficially termed as the Arjunas and are a loosely resonant family of small 
     NEOs which form the near-Earth asteroid belt. It is difficult to imagine that all these objects were formed as they are today within 
     the main asteroid belt and eventually found their way to the NEO population as they were. None the less, the Hungarias have been 
     suggested as a possible direct source for this population (Galiazzo \& Schwarz 2014). NEO orbit models like the one used here show that 
     this is possible, but it is unclear whether the known delivery routes are efficient enough to explain the size of the current 
     population of small NEOs, and in particular the significant number of Arjunas like 1991~VG. Fragments can also be produced {\it in 
     situ} (i.e. in the neighbourhood of the path of the Earth) by multiple mechanisms: subcatastrophic impacts (see e.g. Durda et al. 
     2007), tidal disruptions after close encounters with planets (see e.g. Schunov{\'a} et al. 2014) or the action of the 
     Yarkovsky--O'Keefe--Radzievskii--Paddack (YORP) mechanism (see e.g. Bottke et al. 2006). These processes can generate dynamically 
     coherent groups of genetically related objects ---although YORP spin-up is considered dominant (see e.g. Jacobson et al. 2016)--- which 
     may randomize their orbits in a relatively short time-scale as they move through an intrinsically chaotic environment (see 
     Fig.~\ref{disper}). In addition, the superposition of mean motion and secular resonances creates dynamical families of physically 
     unrelated objects (see e.g. de la Fuente Marcos \& de la Fuente Marcos 2016c) which intertwine with true families resulting from 
     fragmentation. On a more practical side, all these objects are easily accessible targets for any planned NEO sample-return missions 
     (see e.g. Garc{\'{\i}}a Y{\'a}rnoz et al. 2013) or even commercial mining (see e.g. Lewis 1996; Stacey \& Connors 2009). Bolin et al. 
     (2014) have predicted that, while objects like 1991~VG or 2006~RH$_{120}$ are extremely challenging to discover using the currently 
     available telescope systems, the Large Synoptic Survey Telescope or LSST (see e.g. Chesley et al. 2009) may be able to start 
     discovering them on a monthly basis when in full operation, commencing in January 2022.

     Our independent assessment of the current dynamical status and short-term orbital evolution of 1991~VG leads us to arrive to the same
     basic conclusion reached by Brasser \& Wiegert (2008): asteroid 1991~VG had to have its origin on a low-inclination Amor- or 
     Apollo-class object. However and given its size, it must be a fragment of a larger object and as such it may have been produced {\it in 
     situ}, i.e. within the orbital neighbourhood of the Earth--Moon system, during the relatively recent past (perhaps a few kyr ago). 

  \section{Conclusions}
     In this paper, we have studied the dynamical evolution of 1991~VG, an interesting and controversial NEO. This investigation has been 
     carried out using $N$-body simulations and statistical analyses. Our conclusions can be summarized as follows.
     \begin{enumerate}[(i)]
        \item Asteroid 1991~VG currently moves in a somewhat Earth-like orbit, but it is not an Earth's co-orbital now. It has been a 
              transient co-orbital of the horseshoe type in the past and it will return as such in the future. 
        \item Extensive $N$-body simulations confirm that the orbit of 1991~VG is chaotic on time-scales longer than a few decades. 
        \item Our calculations confirm that 1991~VG was a natural satellite of our planet for about one month in 1992 and show that this 
              situation may have repeated multiple times in the past and it is expected to happen again in the future. Being a recurrent
              ephemeral natural satellite of the Earth is certainly unusual, but a few other known NEOs exhibit this behaviour as well.
        \item A realistic NEO orbit model shows that although quite improbable, the presence of objects moving in 1991~VG-like orbits is
              not impossible within the framework defined by our current understanding of how minor bodies are delivered from the main
              asteroid belt to the NEO population. 
        \item Consistently, we find three other minor bodies ---2001~GP$_{2}$, 2008~UA$_{202}$ and 2014~WA$_{366}$--- that move in orbits 
              similar to that of 1991~VG. 
        \item NEOs, moving in 1991~VG-like orbits have a probability close to 0.004 of becoming transient irregular natural satellites of 
              our planet.
        \item Our results show that, although featuring unusual orbital properties and dynamics, there is no compelling reason to consider 
              that 1991~VG could be a relic of human space exploration and definitely it is not an alien artefact or probe.
     \end{enumerate}
     The remarkable object 1991~VG used to be considered mysterious and puzzling, but the new data cast serious doubt on any possible 
     origin for this NEO other than a natural one. We find no evidence whatsoever of an extraterrestrial or intelligent origin for this 
     object. Spectroscopic studies during its next perigee on 2018 February may be able to provide better constraints about its most 
     plausible source, in particular whether it is a recent fragment or not.

  \section*{Acknowledgements}
     We thank the referee, M. Granvik, for his constructive, thorough and very helpful reports, S.~J. Aarseth for providing the code used in 
     this research, A.~I. G\'omez de Castro, I. Lizasoain and L. Hern\'andez Y\'a\~nez of the Universidad Complutense de Madrid (UCM) for 
     providing access to computing facilities. This work was partially supported by the Spanish `Ministerio de Econom\'{\i}a y 
     Competitividad' (MINECO) under grant ESP2014-54243-R. Part of the calculations and the data analysis were completed on the EOLO cluster 
     of the UCM, and we thank S. Cano Als\'ua for his help during this stage. EOLO, the HPC of Climate Change of the International Campus of 
     Excellence of Moncloa, is funded by the MECD and MICINN. This is a contribution to the CEI Moncloa. In preparation of this paper, we 
     made use of the NASA Astrophysics Data System, the ASTRO-PH e-print server, and the MPC data server.

  \bsp
  \label{lastpage}
\end{document}